# *Confidence Assertions in Cyber-Security for an Information-Sharing Environment*

August 2017

Paul B. Kantor, Dennis E. Egan,
Jonathan Bullinger,
Katie McKeon, James Wojtowicz

The preparation of this report was supported by the United States Department of Homeland Security under grant 2009-ST-061-CCI002-08 to Rutgers University. Any opinions or conclusions in this report are those of the authors and do not necessarily represent the positions of CCICADA or of the US Department of Homeland Security.

(This page intentionally blank)

# Confidence Assertions in Cyber-Security for an Information-Sharing Environment

## Contents

**Table of Tables**

**Table of Figures**

**Table of Exhibits**

# Confidence Assertions in Cyber-Security for an Information-Sharing Environment

## 1. Executive Summary

Information sharing is vital in resisting cyberattacks, and the volume and severity of these attacks is increasing very rapidly. Therefore responders must triage incoming warnings in deciding how to act. This study asked a very specific question: "how can the addition of confidence information to alerts and warnings improve overall resistance to cyberattacks." We sought, in particular, to identify current practices, and if possible, to identify some "best practices." The research involved literature review and interviews with subject matter experts at every level from system administrators to persons who develop broad principles of policy. An innovative Modified Online Delphi Panel technique was used to elicit judgments and recommendations from experts who were able to speak with each other and vote anonymously to rank proposed practices.

The following table summarizes a few of the very specific practices that were found to have strong support in our interviews and Delphi panel(s).

Selected Strongly Recommended Practices [See Section 6.1]

| |
|---|
| ASSTN_4: Take a (potentially) hacked machine off-line and keep it turned on for forensic analysis |
| ASSTN_7: Check with your sysadmins, before asserting a threat |
| ASSTN_8: Check with your individual monitoring console before asserting a threat |
| ACQRG_1: Act first on threat reports that come from an organization on our "trusted" list. |

We found agreement that, because of the speed at which cyber threats develop and propagate, it will be good to have automated processes to respond faster than humans can to spreading threats. As we move to the Internet of Things there are, unfortunately, many vulnerable systems that may be "too dumb" to be able to report on their health. For these we may have to rely on SCADA monitoring systems to detect threats, extract some kind of signature and disseminate the warnings (Zhang et al., 2011).

We found that there is no established set of practices for encoding confidence today. However, virtually everyone we interviewed and those participating in our online Delphi session thought that having a standard confidence coding scheme would greatly improve cyber information sharing. Although our results must be regarded as preliminary, there are signs of some consensus among experts around certain practices for asserting and acquiring confidence in cybersecurity reports, as well as for certain policy-level cybersecurity confidence initiatives.

> *I checked an entry or so in our TAXII feed and did not see a field labeled "confidence level" or anything like that. When we spoke with [State Fusion Center], they indicated that they wanted us to check our logs because if we found the same info, it would increase their confidence level. Thus I infer that organizations able to consume the feed in an automated fashion would be able to notice that [we] are reporting hits that were also reported by [the State].*
>
> *A system administrator*

This research also suggests that whatever principles are developed to guide the reporting of confidence in cyber-security incidents should recognize that those reports will eventually be used in automated compilation, either for dissemination or action. It is strongly suggested that those experts who will design the compilation or aggregation be included in discussions about defining the practices for asserting and acquiring confidence. It is also suggested that every report indicate clearly whether it is an initiating report of an incident, or a descendant or secondary report about some incident already reported. Finally, it is suggested that the problem of reporting and using provenance be added to the technical research agenda of the ISE program.

The Modified Online Delphi Process, developed and piloted in this study can be a useful tool for assembling and weighing expert insights at several operational and policy levels. All of these levels should be addressed in parallel to ensure that policies are feasible, and that insights from the front lines guide the development of those policies. In conclusion, time is running out to develop the needed practices for asserting and acquiring confidence in shared reports, and a strong centrally-led effort must complement the collection of good ideas and innovations from operational personnel across the country.

## 2. The Problem

Maintaining cyber security is a constant struggle of limited resources against an ever-expanding world of threats. Information sharing is key to maximizing the effectiveness of those limited resources. The primary form of information sharing is the distribution of reports about new attacks, about how to detect them, and about how to counter them.

> *"Confidence levels are needed in some fashion so that consumers of the information have a level of assurance that the information has been vetted and verified by experts."*
>
> *Government Specialist*

If resources were adequate, every organization could respond to every warning message, taking the steps necessary to counter the new threat. For small organizations this has never been possible. As the flow of threat messages increases, it also becomes impossible for large organizations. Consequently, every organization must develop rules for prioritizing its actions, on the basis of information received. This prioritization considers three characteristics of the threat: the seriousness of its consequences; the costs of taking action against it; and the organization's confidence in the reports about the threat.

We have considered the particular problems associated with the third characteristic of a message about threats: the confidence in cyber threat reports. Our research suggests that there are three important aspects to confidence: deciding how to assert confidence; deciding how to acquire or build up confidence on the basis of reports that are received; and the long-term problem of automating one or both of those steps.

*Asserting* confidence is something that ought to be done by any entity that reports a threat. The report may be quite complex, and the several parts of the report may have differing levels of confidence. Our research suggests that current practice has not reached the stage of being able to assert different levels of confidence in a uniform way. Instead, current discussion centers on whether and how to assert confidence when reporting an incident.

The second aspect that we discuss is the problem of *acquiring* confidence. As one or more reports are received (from outside the organization) the organization must decide how much confidence it has that these represent a serious threat, and that the actions recommended should be taken.

> *"When I see a DHS bulletin and an FBI bulletin both pinpoint a very specific threat source or vulnerability, I have a high degree of certainty that the information has been vetted and may be acted upon in the best interest of managing risk."*
>
> *Government Specialist*

The third aspect, the automation of one or both of these processes will not be discussed here. Our research indicates that the existing frameworks for reasoning with uncertainty would, at this stage of cyber security information sharing, represent good mathematics chasing inadequate data.

Here is a simple example of the kinds of ambiguity that occur. Suppose that a network of information sharing organizations have agreed to use a "high, middle, low" scale to indicate confidence. And suppose that an organization receives two reports with middle confidence. Should these be considered to have the overall effect of lowering confidence (because each of them did not assert high confidence) or should they increase confidence (because they are both reporting the same threat, albeit not with high confidence)? Our research does not suggest a clear answer to this question. One can imagine a setting in which two reports are generated, for example by people who are looking at exactly the same signature. They might each initially think it is a serious threat. But if the first says "it doesn't look too bad to me" and the second says "it doesn't look too bad to me" clearly the combination pushes the threat level down. On the other hand if they each began by thinking there was no threat at all and each revised their opinion by saying "on the other hand perhaps there really is a problem here" then the combination of them is a more serious warning than either of them alone. An examination of the underlying mathematical issues appears in an Appendix, Section 12.

## 2.1. Cyber Security Responders must prioritize their actions

> *"Looking at the sources is more important than looking at the data (which can be manipulated easily). Must trust the originating organization. Most trust is built up over time … Trust is earned and confidence is based on that."*
>
> *Policy-level expert*

Having received multiple alerts or warnings a system administrator or operator must decide what to do first, and what to not do at all. This decision will always reflect the decision-maker's capabilities, the resources available, and the estimated harm if the action is delayed or not taken at all ("seriousness"). Some actions can be completed promptly, others require considerable lead time. Some can be done without even being noticed by system users; others may require taking systems down, or imposing new procedures on all the users. All of these are "costs." But the consideration most important for the present analysis is not any of these operational factors. It is the believability of the alert or warning itself. We do not speculate here on whether that factor is considered first, or second, by a decision maker, and so the results of this study cannot provide a guide for which action to take. An example is shown in Table 2-1.

Table 2-1. Contrasting Priority Rules

| Action | Seriousness | Confidence | Cost(Time) of Action |
|--------|-------------|------------|----------------------|
| Patch A | High | High | High |
| Patch B | High | Medium | Low |

For the possible actions shown in Table 1, let us suppose that both are "high seriousness of consequences" that is, they protect against serious threats. We suppose that the cause for each action is an alert, and that one alert has high confidence and the other has only medium confidence. With no other considerations, it seems clear that Patch A should be done first.

However, a specific threat report may call for a centralized response, such as an update to the operating system, or a more detailed specific response, such as to apply a specific patch to some software throughout the organization. Suppose we also know that Patch B can be done quickly (that is, has Low cost (measured in time)) while Patch A requires separate visits to each of five departments. From this perspective Patch B should be done first. We cannot make any recommendation as to which principle a security administrator should apply, and the study of current practices in this regard is outside the scope of this project.

## 3. Methods

We used multiple methods to build a set of candidate practices, and then to work towards a procedure for ranking them, using the aggregated judgment of experts with operational responsibility for maintaining the integrity of cyber systems.

### 3.1. Interviews with Experts

Interviews of subject matter experts are a key aspect of our research. Experts were identified from previous information sharing projects conducted by CCICADA for the U.S. Coast Guard (USCG) and for the Office of the Director of National Intelligence (ODNI). We pursued a snowball approach for recruiting additional interviewees by asking each expert, at the close of our interview, to suggest the names of additional people we could contact. The interviewees included cybersecurity executives and managers, as well as those working at an operational level. We also spoke to officers of two large standards organizations that develop and host information sharing tools and protocols.

> *"There is a false sense of precision which is true of most confidence rating schemes when based on a single analyst's opinion. You need common, agreed terminology and methodologies to assign."*
>
> *Policy-level expert*

Interviews were conducted by telephone, typically with two researchers and the subject matter expert on a conference bridge. The researchers took notes during the interview, which were transcribed, edited, and posted to a project archive site for sharing with the project team. The interviews were not for attribution and typically lasted between 30 to 60 minutes.

These open-ended interviews were designed to elicit discussion from the expert. After some brief introductory remarks, the semi-structured interview was guided by the following questions:

Q1. Do you have some experience in assigning confidence to incident reports? As an originator or as a recipient? Can you tell me something about how that works?

Q2. Do you think confidence information should be a required part of every incident report? Can you tell me why? Or why not?

> *"There is an SOP for incident response. They go to the machine; take it off line. Then they do a scan and try several removal tools. In the most recent incident, their tools could not remove the problem. They don't examine the code of the problem. If they can remediate they do. If not, they wipe the machine and reboot it."*
>
> *University Sysadmin*

Q3. Above and beyond confidence that is formally asserted in an incident report, do you have any "rules of thumb" that you use in deciding how much weight to give to the report?

Q4. In your personal experience is confidence expressed in numbers? By a color scale? In words? Which works well? Why?

Q5. Now that you have a clearer idea of what we are after here, can you suggest one or two other people whom we could call, to learn more? May we use your name when we do?

Q6. Is there something else that I should have asked you, that I didn't?

Eight experts were interviewed using this guide. Additional information was obtained in less formal conversations and email exchanges with another four cybersecurity experts from federal government organizations and the private sector.

## 3.2. Literature Reviews

In parallel with the subject matter expert interviews, we conducted an extensive literature review looking for information about how confidence in uncertain situations is communicated in other sectors. All roads seemed to lead towards the intelligence community, and those findings are summarized in Section 4. We also reviewed a number of efforts that are directed towards information sharing, to look for models of how confidence is, or might be, expressed. Those efforts are summarized in an Appendix, Section 10.

> *"When a person suspects that the computer he uses (they are all "thin clients") is compromised the machine is left with power on, taken off the network, and replaced. A forensic team then works with the unit to learn what it can about the attack."*
>
> *DOD Sysadmin*

## 3.3. Formulation of Proposed Practices

The project team carefully reviewed the literature summaries and interview notes, and worked to formulate very specific practices. It proved challenging to move beyond general principles. Such principles may be interpreted in so many different ways that one could not be sure that all who agreed held the same idea in their minds when agreeing. By considerable critical editing, we eventually developed a collection (see Section 5, below) of candidate practices.

### 3.4. Modified Online Delphi Process

**Participants.** The original work plan called for holding a number of focus group meetings, in conjunction with national gatherings of experts in cyber-security, particularly in the Washington D.C. area. However, as our interviews proceeded, as described above, we were advised that the experts who assemble in and around Washington represent one particular slice of the cybersecurity community. In particular, this subgroup were described as being likely to advocate for some particular positions. At the same time, they are somewhat removed from the day-to-day operational setting, for which good practices are most urgently needed. With this guidance, we extended our snowball recruitment to identify experts who are closer to the "front lines:" system operators and administrators, and people working in fusion centers where threat messages are received and disseminated.

**Method.** With the practices and panelists defined, we considered how to move beyond discussion to produce more concrete ranking or scoring of the proposals. A well-established method for producing such rankings is the Delphi process, invented many years ago at RAND (Dalkey and Helmer, 1963), and extensively studied in the social sciences. Essentially the Delphi process permits two or more rounds of estimation and revision, while eliminating the group process phenomena that can lead a small group to a "false consensus" that does not represent the best judgment of each of the participants. However, classical Delphi takes considerable time, as the experts must submit their judgments, those judgments must be aggregated, and the aggregate information (specifically *not* labeled as to who proposes which assessment) must be redistributed.

> *"We use the following criteria for our auto-generated Indicator of Compromise (IOCs):* **High***: Confirmation by analysis or confirmation/detection by multiple sources.* **Medium***: Detection by single source or logical association* **Low***: Suspicious activity or reconnaissance*
> *I would love to see a more inclusive rating with threat, reputation and deprecation/expiration combined."*
> *State Cyber Expert*

To accelerate the process we devised a family of online spreadsheets, using Google Sheets®. Once the principles were defined, we introduced one further layer of abstraction, creating an automated process (using Google Scripts) so that researchers may easily build the needed number of input sheets, and the aggregation or master sheets. These are described more fully in an Appendix (Section 11), and will be available from CCICADA, under a licensing agreement. Examples of the Expert input sheet, and the Master or Aggregator sheet are shown in Figure 1. Panelist Sheet for Delphi Process (excerpt). and Figure 2. Master Sheet for Delphi Process (excerpt).

File  Edit  View  Insert  Format  Data  Tools  Add-ons  Help   Last edit was made on June 19 by Confidence Account

Instructions

## Instructions

We would like you to evaluate each of the practices below. Specifically, we would like you to consider the criteria on the right and select all that apply by placing an X in the cells below.

| | |
|---|---|
| H  Have Tried it | Do you have personal experience with this practice? |
| F  Feasible | Is the practice realistic and possible to implement |
| S  Sustainable | Can the practice be sustained indefinitely? |
| I  Important | |
| N  Not Important | |

You may leave a comment for each practice which will be sent (anonymously) with your evaluation

| Practices | | | | | | | Your Evaluation |
|---|---|---|---|---|---|---|---|
| Group | Practice | H | F | S | I | N | Comments |
| Asserting | ASSTN_1: Send out a "low confidence" report concerning any device that seems to be hacked | | X | | | X | Creates a lot of noise |
| Asserting | ASSTN_2: Handle desktop problems remotely, and decide whether to issue a report | X | X | X | | | |
| Asserting | ASSTN_3: Reimage a (potentially) hacked machine, and issue no report | X | X | | | X | Good for endpoints that are subject to frequent comprimise if you know there |
| Asserting | ASSTN_4: Take a (potentially) hacked machine off-line and keep it turned on for forensic analysis | X | X | X | X | | Necessary for anything high-value or high efficacy |
| Asserting | ASSTN_5: Check with VTAPI and only report bad URLs that are on the list | | | | X | | |
| Asserting | ASSTN_6: Check outside resources before asserting a threat | | | | | X | Not sure all that filtering is necessary, seems the velocity vector is important t |
| Asserting | ASSTN_7: Check with your sysadmins, before asserting a threat | X | X | X | | | |
| Asserting | ASSTN_8: Check with your individual monitoring console before asserting a threat | X | X | X | X | | |
| Acquiring | ACQRG_1: Act first on threat reports that come from an organization on our "trusted" list. | X | X | X | X | | |

Figure 1. Panelist Sheet for Delphi Process (excerpt).

## Facilitator Instructions

Your sheet is automatically updated with expert input. In order to export the results to the experts, you will need to find 'Round One' on the menu bar above and click 'Tally Results.' You may re-tally at any time.

| Practices | | Expert A | | | | | | Expert B | | | | | | Expert C | | | | | | |
|---|---|---|---|---|---|---|---|---|---|---|---|---|---|---|---|---|---|---|---|---|
| Group | Practice | H | F | S | I | N | Comments | H | F | S | I | N | Comments | H | F | S | I | N | Comments | H F |
| Asserting | ASSTN_1: Send out a "low confidence" report concerning any device that seems to be hacked | | X | | | X | Creates a lot of r | | | | | | | x | x | | | x | Early warning to | |
| Asserting | ASSTN_2: Handle desktop problems remotely, and decide whether to issue a report | X | X | X | | | | | | | | | | x | x | x | x | | | |
| Asserting | ASSTN_3: Reimage a (potentially) hacked machine, and issue no report | X | X | | | X | Good for endpoints that are subject to frequent comprimise if you know there is not much value | | | | | | | x | x | | | x | I think this is not a goox | |
| Asserting | ASSTN_4: Take a (potentially) hacked machine off-line and keep it turned on for forensic analysis | X | X | X | X | | Necessary for anything high-value or high efficacy | | | | | | | x | x | x | x | | | x |
| Asserting | ASSTN_5: Check with VTAPI and only report bad URLs that are on the list | | | | X | | | | | | | | | | | | | | We use multiple source x | |
| Asserting | ASSTN_6: Check outside resources before asserting a threat | | | | | X | Not sure all that filtering is necessary, seems the velocity vector | | | | | | | x | x | x | x | | We always check with x | |
| Asserting | ASSTN_7: Check with your sysadmins, before asserting a threat | X | X | X | | | | | | | | | | x | x | x | x | | In almost all cases we x | |
| Asserting | ASSTN_8: Check with your individual monitoring console before asserting a threat | X | X | X | X | | | | | | | | | x | x | x | x | | We find high value in o x | |
| Acquiring | ACQRG_1: Act first on threat reports that come from an organization on our "trusted" list. | X | X | X | X | | | | | | | | | x | x | x | x | | Everything must be pri x | |
| Acquiring | ACQRG_2: ONLY consider the sources in "acquiring confidence." | | | | | | | | | | | | | | | | | | We consider verificatio x | |
| Acquiring | ACQRG_3: Always trust the Emerging Threats List. | | | | | | | | | | | Blind trust is another thing that can be abused - need to be careful | | | | | | Additional validation is x | |
| Acquiring | ACQRG_4: Prioritize reports that have a valid expiration date | X | | | | | | | | | | | | | | | | | | |
| Acquiring | ACQRG_5: Do not act on a report that seems to be only one analyst's opinion | | | X | | | While sustainable, this is not a good idea - sometimes a single a x | | | | | | | x | x | | | | One analyst could be p x | |
| Acquiring | ACQRG_6: Act first on threat reports that clearly have had human | | | | | | | | | | | | | | | | | | | |

Figure 2. Master Sheet for Delphi Process (excerpt)

The Modified Online Delphi Process involved three team members: the facilitator, who read the instructions to the participants, and led the discussions at each stage; the designer, who ensured that the software was behaving as intended, and dealt with any user behavior that had not been anticipated, and the annotator, who recorded key points that were not entered into the Delphi Response Sheets. Findings are summarized in Section 6.

# 4. Literature Review

## 4.1. Summary Findings

An extensive discussion of the literature related to this topic appears in an Appendix, Section 10. The several schemes for converting estimated probabilities to words are briefly summarized in this section.

It has been known for some time that the intelligence community seeks to develop regular and accepted principles for asserting confidence in its estimates (Kent, 1963). Even though that report was written 54 years ago, it appears that, in applying these ideas to cyber-security, the present effort is still well ahead of the curve.

Several specific scales came up in our research. The most detailed found in our interviews is in an unclassified excerpt provided by the FBI, Exhibit 4-1. A more detailed exposition subsequently became available in ODNI (2017) as excerpted in Exhibit 4-2. An interesting discussion is given in Section 10.4.2, from the US Coast Guard. We summarize various schemes for relating terminology to subjective estimates of probability in Table *4-1*

Exhibit 4-1 FBI Estimative Terminology

| UNCLASSIFIED | | | | | | | |
|---|---|---|---|---|---|---|---|
| *Terms of Likelihood* | Almost No Chance | Very Unlikely | Unlikely | Roughly Even Chance | Likely | Very Likely | Almost Certain(ly) |
| *Terms of Probability* | Remote | Highly Improbable | Improbable (Improbably) | Roughly Even Odds | Probable (Probably) | Highly Probable | Nearly Certain |
| | 1-5% | 5-20% | 20-45% | 45-55% | 55-80% | 80-95% | 95-99% |

(U) Unless otherwise stated, the FBI does not derive judgments via statistical analysis. **(U) Expressions of Likelihood (or Probability)**

(U) Phrases such as "the FBI judges" and "the FBI assesses," and terms such as "likely" and "probably" convey analytical judgments and assessments. The chart approximates how expressions of likelihood and probability correlate with percentages of chance.

## ESTIMATIVE LANGUAGE

Estimative language consists of two elements: judgments about the likelihood of developments or events occurring and levels of confidence in the sources and analytic reasoning supporting the judgments. Judgments are not intended to imply that we have proof that shows something to be a fact. Assessments are based on collected information, which is often incomplete or fragmentary, as well as logic, argumentation, and precedents.

**Judgments of Likelihood.** The chart below approximates how judgments of likelihood correlate with percentages. Unless otherwise stated, the Intelligence Community's judgments are not derived via statistical analysis. Phrases such as "we judge" and "we assess"—and terms such as "probable" and "likely"—convey analytical assessments.

*Percent*

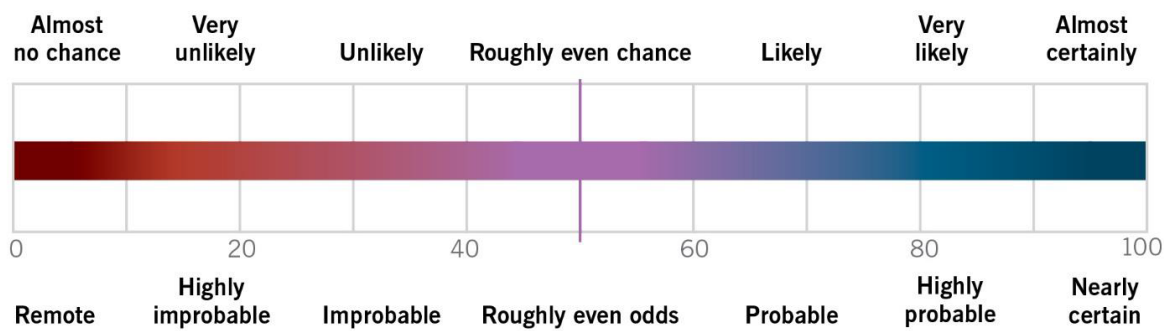

**Confidence in the Sources Supporting Judgments.** Confidence levels provide assessments of the quality and quantity of the source information that supports judgments. Consequently, we ascribe high, moderate, or low levels of confidence to assessments:

- **High confidence** generally indicates that judgments are based on high-quality information from multiple sources. High confidence in a judgment does not imply that the assessment is a fact or a certainty; such judgments might be wrong.

- **Moderate confidence** generally means that the information is credibly sourced and plausible but not of sufficient quality or corroborated sufficiently to warrant a higher level of confidence.

- **Low confidence** generally means that the information's credibility and/or plausibility is uncertain, that the information is too fragmented or poorly corroborated to make solid analytic inferences, or that reliability of the sources is questionable.

The following table summarizes the several schemes that are used to map numerical estimates of probability into terms from the common language of uncertainty. There is consistency within the United Stated, but not, apparently, with the British practices.

Table 4-1. Terminology Related to Probability

| Percentage | Recommended terms or codes | | | | |
|---|---|---|---|---|---|
| | Clapper Directive ODNI | USCG | FBI | Admiralty Code (Terms) | Admiralty Code (Numbers) |
| 01-05% | Remote | Remote | Remote | Improbable | **5** |
| 05-20% | Highly Improbable | Highly Improbable | Highly Improbable | Doubtfully True | **4** |
| 20-45% | Improbable (Improbably) | Improbable (Improbably) | Improbable (Improbably) | | |
| 45-55% | Roughly even odds | Roughly even odds | Roughly even odds | | |
| 55-80% | Probable (Probably) | Probable (Probably) | Probable (Probably) | Possibly True | **3** |
| 80-95% | Highly Probable | Highly Probable | Highly Probable | Probably True | **2** |
| 95-99% | Nearly Certain | Nearly Certain | Nearly Certain | Confirmed | **1** |

**4.2.Discussion**

Aggregating information provided with such broad ranges as are shown in Table *4-1* would clearly present its own challenges. However, our research reveals that in the cyber-security community, even these practices have not yet taken root. Consistent and widely used practices are yet to be defined. This problem does contain a hidden opportunity. Perhaps the long-term problems of automated reasoning with cyber security reports will be more tractable, if reporting practices are developed with an eye towards reasoning with diverse and possibly conflicting information of varying (un)certainty

# 5. Specific Practices Developed

Specific practices might be organized in several ways. We originally thought in terms of a relatively abstract characterization, in which a report might be seen to have assertions of various *kinds*, about the threat, about its relation to other known threats (the "campaign") and about the provenance of the report. However, that perspective did not match well with the specific practices that emerged in our research.

> *"Our partners ask: 'does high (threat level) mean I should act at once, or look for it and get back to you?' So there is room for further clarification of the recommended action, as well as the priority."*
>
> *State Cyber Expert*

The specific practices that we found seemed, instead, to organize themselves into three broad categories: Asserting Confidence, Acquiring Confidence, and Policy.

## 5.1. Asserting Confidence

The first group of practices deals with the problem of asserting confidence, which we operationalized for the respondents as: *if you suspect you have been attacked, your suspicions may turn into confidence, or remain somewhat uncertain. However that may be, when you send a warning out (either within your organization, or out to an information sharing center) you have some level of confidence about what you are reporting. And so some of the practices listed here most likely apply when you are deciding how confident you are in what you are reporting. We can call these aspects "asserting confidence."* These practices are shown in Table 5-1

Table 5-1. Practices for Asserting Confidence

| Practice |
| --- |
| ASSTN_1: Send out a "low confidence" report concerning any device that seems to be hacked |
| ASSTN_2: Handle desktop problems remotely, and decide whether to issue a report |
| ASSTN_3: Reimage a (potentially) hacked machine, and issue no report |
| ASSTN_4: Take a (potentially) hacked machine off-line and keep it turned on for forensic analysis |
| ASSTN_5: Check with VTAPI and only report bad URLs that are on the list |
| ASSTN_6: Check outside resources before asserting a threat |
| ASSTN_7: Check with your sysadmins, before asserting a threat |
| ASSTN_8: Check with your individual monitoring console before asserting a threat |

## 5.2. Acquiring Confidence

The second group of practices deals with the problem of acquiring confidence, which we operationalized for the respondents as: *Second, you also receive warnings and threat information. You have to decide what action(s) to take. And that decision is likely to be affected by your confidence about the warning. That confidence might come in many ways: the report may contain* *a statement about its own confidence in itself. It might come from a source that you trust highly (or perhaps, not very highly). It might be that you are getting the same reports from multiple sources, and that increases your confidence in all of them. Let's call these aspects "acquiring confidence."* These practices are shown in Table 5-2.

Table 5-2. Practices for Acquiring Confidence

| |
|---|
| ACQRG_1: Act first on threat reports that come from an organization on our "trusted" list. |
| ACQRG_2: ONLY consider the sources in "acquiring confidence." |
| ACQRG_3: Always trust the Emerging Threats List. |
| ACQRG_4: Prioritize reports that have a valid expiration date |
| ACQRG_5: Do not act on a report that seems to be only one analyst's opinion |
| ACQRG_6: Act first on reports that clearly have had human vetting |
| ACQRG_7: Act first on reports that clearly are generated by computer monitoring |
| ACQRG_8: Act with complete confidence if a report comes from a government entity |
| ACQRG_9: Base our confidence on knowing the human who made the report |
| ACQRG_10: Base our confidence on your history with the source |
| ACQRG_11: We have higher confidence in reports that have more detail |
| ACQRG_12: We have higher confidence in reports that have an expiration date |
| ACQRG_13: We have higher confidence in reports that are more recent |

**5.3. Policies for Confidence**

The third group of practices are not related to current day-to-day operations, but represent *policies* whose implementation would advance the overall effort to reason with and about confidence in an information sharing environment. We operationalized these for the respondents as: *"Later, in Round Three, will also ask for your thoughts about some proposed "policy level" changes."* These practices are shown in Table 5-3.

Table 5-3. Practices at the Policy Layer

| |
|---|
| POL_1 DHS should define a standard language for asserting confidence |
| POL_2 Some checklist for event verification should be established for small organizations to use |
| POL_3 The community should develop a "reputation scale" 1 to 100 for entities that report attacks |
| POL_4 Report expiration dates should be less than a year |
| POL_5 STIX should have a way of embedding supporting evidence in the indicator |

# 6. Findings of the Delphi Process

## 6.1. Practices for Asserting or Acquiring

The principal findings of the Delphi session(s) concern practices for asserting or acquiring confidence. They are summarized numerically in Exhibit 6-1. The numerical entries in the cells report the number of panelists (out of a total of 4) who checked the corresponding box for a particular practice.  So, for the practice listed in the first row of the table, 3 panelists had personally tried it (H), all 4 thought it was feasible (F), all 4 thought it was sustainable (S), all 4 thought it was important (I) and none thought it was not important (N). Note that the descriptions are truncated for ease of presentation. They are stated more fully in Table 5-1 and  Table 5-2.

| Practice | H | F | S | I | N |
|---|---|---|---|---|---|
| ASSTN_4: Take a (potentially) hacked machine off-line and keep | 3 | 4 | 4 | 4 | 0 |
| ASSTN_7: Check with your sysadmins, before asserting a | 3 | 4 | 4 | 4 | 0 |
| ASSTN_8: Check with your individual monitoring console before | 3 | 4 | 4 | 4 | 0 |
| ACQRG_1:  Act first on threat reports that come from an | 3 | 4 | 4 | 4 | 0 |
| ASSTN_2: Handle desktop problems remotely, and decide | 3 | 4 | 4 | 3 | 0 |
| ACQRG_10: Base our confidence on your history with the | 2 | 4 | 4 | 3 | 0 |
| ACQRG_13: We have higher confidence in reports that are more | 2 | 4 | 4 | 3 | 0 |
| ASSTN_6: Check outside resources before asserting a threat | 2 | 3 | 3 | 3 | 1 |
| ACQRG_6: Act first on reports that clearly have had human | 2 | 3 | 2 | 3 | 0 |
| ACQRG_11: We have higher confidence in reports that have | 1 | 3 | 3 | 2 | 0 |
| ACQRG_9: Base our confidence on knowing the human who | 1 | 3 | 2 | 2 | 0 |
| ACQRG_7: Act first on reports that clearly are generated by | 1 | 2 | 3 | 2 | 0 |
| ASSTN_1: Send out a "low confidence" report concerning any | 2 | 4 | 2 | 1 | 2 |
| ACQRG_5: Do not act on a report that seems to be only one | 2 | 3 | 3 | 1 | 0 |
| ACQRG_12: We have higher confidence in reports that have an | 1 | 3 | 3 | 1 | 1 |
| ASSTN_5: Check with VTAPI and only report bad URLs that are | 0 | 1 | 1 | 1 | 1 |
| ASSTN_3: Reimage a (potentially) hacked machine, and issue | 2 | 3 | 3 | 0 | 2 |
| ACQRG_4: Prioritize reports that have a valid expiration date | 1 | 3 | 2 | 0 | 1 |
| ACQRG_2: ONLY consider the sources in "acquiring | 1 | 2 | 2 | 0 | 1 |
| ACQRG_8: Act with complete confidence if a report comes from | 2 | 2 | 2 | 0 | 2 |
| ACQRG_3: Always trust the Emerging Threats List. | 0 | 1 | 1 | 0 | 1 |

Exhibit 6-1. Delphi: Assessment of Operational Practices

We see that none of the practices have been used by all of the panelists. Several were judged to be feasible, but not always sustainable, and they have been ranked in order of the number of panelists who judged the practice to be important, then by the number who judged it feasible, and finally by the number who judged it sustainable.

We call attention to the four practices that were scored important, feasible and sustainable by all of the respondents, and include them in our Executive Summary as strongly recommended. They are ASSTN_4,7 and 8, and ACQRG_1, as summarized in Table 6-1.

Table 6-1. Delphi: Selected Strongly Recommended Practices

| |
|---|
| ASSTN_4: Take a (potentially) hacked machine off-line and keep it turned on for forensic analysis |
| ASSTN_7: Check with your sysadmins, before asserting a threat |
| ASSTN_8: Check with your individual monitoring console before asserting a threat |
| ACQRG_1: Act first on threat reports that come from an organization on our "trusted" list. |

There is a second group that were scored as feasible by all panelists, but not scored important by all. With a very small panel such as this we simply note the difference, but do not offer any conclusions. Perhaps of interest is the bottom of the list, revealing that no panelist had blindly trusted the emerging threats list, and no one considered it to be important to do so.

Finally, we present the several comments that panelists provided. During the Delphi process these were not linked to specific people, although in some cases an individual claimed ownership of a comment, during the discussion. Since they may be useful in future replications of this process, they are included here in *Table 6-2*.

Table 6-2. Delphi: Panelist Comments

| | |
|---|---|
| ASSTN_7: Check with your sysadmins, before asserting a threat | *In almost all cases we check with Sysadmins - exception is when they might be involved |
| ASSTN_8: Check with your individual monitoring console before asserting a threat | *We find high value in our console - it validates the data we review |
| ACQRG_1: Act first on threat reports that come from an organization on our "trusted" list. | *Everything must be prioritized - working with data from trusted sources is first - then additional information is considered |
| ACQRG_10: Base our confidence on your history with the source | *There are good experiences with some and not with others - this should be considered strongly |
| ASSTN_6: Check outside resources before asserting a threat | *We always check with external resources - they are constantly updating |
| ACQRG_6: Act first on reports that clearly have had human vetting | *Non-vetted information is valuable, but vetted is more actionable |
| ACQRG_11: We have higher confidence in reports that have more detail | *Top level data without substantiation is less valuable - we need detail to examine when we can |
| ACQRG_9: Base our confidence on knowing the human who made the report | *As mentioned above, we often count on high value analysts |
| ACQRG_7: Act first on reports that clearly are generated by computer monitoring | *Near real time is important - attaining information directly from the networks we're continuously monitoring is valuable and actionable |
| ASSTN_1: Send out a "low confidence" report concerning any device that seems to be hacked | *Early warning to our community for a suspected device is important<br>*If sending a low confidence report, it is very important to make it clear that the information is low confidence and requires additional validation by the recipient. |

| ACQRG_5: Do not act on a report that seems to be only one analyst's opinion | *One analyst could be proven to be highly reliable and accurate - there is a risk, but a low one |
|---|---|
| ASSTN_5: Check with VTAPI and only report bad URLs that are on the list | *We use multiple sources for blacklisting IPs |
| ASSTN_3: Reimage a (potentially) hacked machine, and issue no report | *I think this is not a good practice and wastes valuable insights. Reimaging after data collection makes sense |

## 6.2. Policy Practices

In a third round of discussion we examined the set of proposed policy recommendations. The results are summarized in **Error! Reference source not found.**.[1]

Table 6-3. Delphi: Assessment of Policy Practices

| Practices | | Tallied Judgements | | | |
|---|---|---|---|---|---|
| Group | Practice | H | F | I | N |
| Policy Level Practices | POL_1 DHS should define a standard language for asserting confidence | 0 | 3 | 3 | 0 |
| Policy Level Practices | POL_2 Some checklist for event verification should be established for small organizations to use | 0 | 3 | 3 | 0 |
| Policy Level Practices | POL_3 The community should develop a "reputation scale" 1 to 100 for entities that report attacks | 0 | 3 | 3 | 0 |
| Policy Level Practices | POL_4 Report expiration dates should be less than a year | 0 | 2 | 2 | 0 |
| Policy Level Practices | POL_5 STIX should have a way of embedding supporting evidence in the indicator | 0 | 3 | 3 | 0 |

We see uniform support for the policies, with the exception of the notion that there should be a specific expiration date of less than one year. As the Delphi process proceeded (see Section 6.3), it emerged that the preferred idea was the somewhat less rigid notion that one should "create a freshness element to inform the user of its current relevance and enable automated tools to sort properly." This captures the idea that recency is important, without freezing the time period artificially at one year.

## 6.3. Additional Suggested Practices

The Modified Online Delphi Process allowed for participants to propose additional practices, and they did. These practices, together with their scores, are shown in *Table 6-4*.

Table 6-4. Delphi: Additional Suggested Practices

| Practice Descriptions | H | F | S | I | N |
|---|---|---|---|---|---|
| Add source confidence levels as metadata to cyber threat intel recorded in intel management system | 0 | 2 | 3 | 3 | |

---

[1] Note that one of the panelists had to leave the process, to attend to an operational emergency.

| | | | | |
|---|---|---|---|---|
| Compare multiple sources to raise confidence | 1 | 2 | 3 | 3 |
| compile multiple sources in a dashboard to compare and collate to support decision making | 2 | 3 | 2 | 1 (*) |
| create a freshness element to inform the user of its current relevance and enable automated tools to sort properly | 2 | 3 | 3 | 0 |
| Create a system that will score/summarize threats with some level of consistency | 2 | 3 | 3 | 0 |
| document threat analysis processes to reduce time to decision and increase response effectiveness | 2 | 3 | 3 | 0 |
| enusre there is sufficient content and relevance for intelligence to make it actionable | 2 | 3 | 3 | 0 |
| store or remove outdated intelligence to reduce data to be analyzed | 2 | 3 | 3 | 0 |
| Updated rating on issues as they are addressed (partially or completely) in the field | 2 | 3 | 3 | 0 |
| Velocity and domain - how quickly are threats moving and likely / verified industry targets | 1 | 2 | 3 | 0 |

*One individual considered the proposal for a dashboard to be unnecessary, explaining that the practice of comparing multiple sources subsumes the specific notion of a blackboard, and makes it unnecessary.

With that single exception there was complete unanimity that all of the proposed practices are important. None have been used by more than one respondent. We are not quite sure how to interpret the fact that all considered the practices to be sustainable, but only two panelists considered them feasible. A possible explanation is that they are considered "sustainable once they are in place, but that it is not yet feasible to put them in place." This anomaly merits further study.

# 7. Findings and Conclusions

The key findings of this report are the specific practices described in Section 5, and the available numerical results of the Delphi process. In this process we sought categorical judgments, and the results are summarized in the tables of Section 6.

All of the experts interviewed, whether one on one, or in a Delphi panel, agreed that it is vital to be able to assert confidence when sharing information about cyber threats and specific incidents. They also agreed on the importance of being able to use incoming reports to acquire enough confidence to know what to do first, and what to do next. There was, in addition, agreement that there need to be specific practices (as a opposed to very high level statements of intent) so that operational personnel know what to do, and know how to do it.

Although there is a substantial literature on the problem of asserting confidence, there are no emergent practices for doing so. It appears that operational personnel gain confidence when threat information comes from a known trusted source, or is corroborated as coming from multiple sources, or shows up in multiple ways on their internal monitoring consoles.

## 8. Discussion

This study has addressed an issue that was seen to be important when the study was initiated and, after the cyber attacks of 2016 on the United States electoral process, and the world wide ransomware attacks in early 2017, its importance is even more apparent. There is a danger that, as the problem is addressed by a multitude of uncoordinated agencies, the solutions will become snarled in interagency "turf wars." Many of the documents reviewed showed the results of the "top down" approach to cyber-security, resulting in long tables of principles, and of "progress metrics" which, in most cases, amount to asserting that "something like what the principle calls for is indeed being done."

It may be more effective to institute a parallel, perhaps "crowd-sourced" process to elicit more and better practices, perhaps using those reported here to start the conversation. In spite of the sensible concerns about leaking important national secrets through an overly open process, the benefits of wide dissemination and uptake will likely outweigh those concerns. However, strong leadership will be needed to empower the operational personnel, and to synthesize and promulgate the resulting practices.

We found agreement that, because of the speed at which cyber threats develop and propagate, it will be good to have automated processes to respond faster than humans can, to spreading threats. As we move to the Internet of Things there are, unfortunately, many vulnerable systems that may be "too dumb" to be able to report on their health. For these we may have to rely on SCADA systems to detect threats, extract some kind of signature and disseminate the warnings (Zhang et al., 2011).

In conclusion, time is short to develop the needed practices for asserting and acquiring confidence in shared reports, and a strong centrally-led effort must complement the collection of good ideas and innovations from operational personnel across the country.

# 9. References and Literature Cited

Prepared by Jonathan L. Bullinger

## 10.1.   Review of Estimative Language Formulations

Both the "Clapper Directive" and the FBI guidance provide standards for expressing both likelihood and probability, presented as a table of seven columns and three rows with vocabulary and percentages. Analysts are reminded not to mix these two standards (analyst's confidence in an assessment and the degree of likelihood). The USCG also uses the same terminology without the percentages, expressed as a continuum and visualized as blue arrows. The FBI also uses this terminology visualizing it as a grayscale table with percentages and unlike the other departments, denoting which row is for likelihood and which is for probability. Terms for likelihood include: "almost no chance," "very unlikely," "unlikely," "roughly even chance," "likely," "very likely," and "almost certain(ly)." Terms for probability include: "remote," "highly improbable," "improbable," "roughly even odds," "probable," "highly probable," and "nearly certain." Both the USCG and FBI documents provide definitions for high, middle or moderate, and low confidence. These definitions comprise the following concepts: corroboration, reliability, deception, how critical the assumptions are, and reasoning based on analytical techniques. The presence, absence, or strength of these various components distinguishes the levels between high, medium/moderate, and low. The Admiralty Code, referenced in the US Army FM 2-22.3 (FM 34-52) (2006) uses source reliability and information content as metrics; the former graded from A to F and the latter from 1 to 6. A and 1 refer to Reliable and Confirmed while F and 6 both refer to "cannot be judged." Refer to Section 10.2 for the original tables and definitions for Clapper, USCG, FBI, and Admiralty.

Other sources help provide context for our current task. For example, the State Archives Department, Minnesota Historical Society (2002) document on Trustworthy Information Systems chose the term trustworthy "because it denotes integrity, ability, faith, and confidence" and the records within that system as "reliable and authentic." Other potentially useful vocabulary comes from The U.S. Department of Justice's Global Justice Information Sharing Initiative Information Quality Program Guide (2010) that defines information quality as encompassing "dimensions of accuracy, timeliness, completeness, and security."

Expert elicitation has revealed additional specifics regarding how they formally and informally define confidence. E4 states the scenario, source, and level of detail provided – in particular logs with source / dest. IP and times – increases the confidence to take action. Also alerts coming from within their own team of monitors triggers high confidence. Another expert, E1 looks to the popular ratings app Yelp to convey the concept of a "bunch of ratings" rather than one singular powerful voice. E3 stated that confidence levels from the Intelligence Community have not penetrated much into the common cyber-security industry community. "It is rare that I see true confidence levels. Confidence levels are needed in some fashion so that consumers of the information have a level of assurance that the information has been vetted and verified by experts." E2 stated that confidence is only used with regard to a reputation class and also felt there is a need for an expiration period regarding bad information. E5 said that most of the trust has to come from the originating source. The source is to be trusted, not the information itself. E6, in a joint interview with E5 mentioned that it is easier to assign confidence to something technical (e.g. a specific IP address). However, there is a false sense of precision, which is true of most confidence rating schemes when based on a single analyst's opinion. E4 mentioned there were four sources for reports: outside, inside, sysadmins, and their own consoles. E4 felt that internally, a plain language "FYI" and also starting a work ticket indicates higher confidence to the team. This indicates that they do use their own internal confidence labeling system informally.

## 10.2. Alternative Methods for Expressing Probabilities and/or Confidence

Table 10-1. Comparing Numerical Values and Terminology from Four Sources

| | Clapper Directive | USCG | FBI | Admiralty Code | Admiralty Code (#s) |
|---|---|---|---|---|---|
| **Percentage** | **Terms** | | | | |
| 01-05% | Remote | Remote | Remote | Improbable | 5 |
| 05-20% | Highly Improbable | Highly Improbable | Highly Improbable | Doubtfully True | 4 |
| 20-45% | Improbable (Improbably) | Improbable (Improbably) | Improbable (Improbably) | | |
| 45-55% | Roughly even odds | Roughly even odds | Roughly even odds | | |
| 55-80% | Probable (Probably) | Probable (Probably) | Probable (Probably) | Possibly True | 3 |
| 80-95% | Highly Probable | Highly Probable | Highly Probable | Probably True | 2 |
| 95-99% | Nearly Certain | Nearly Certain | Nearly Certain | Confirmed | 1 |

Table 10-2. Comparing Definitions and Code Letters for Expressions of Confidence from Three Sources:

| Source | | | | |
|---|---|---|---|---|
| Meaning | USCG | FBI | Admiralty Code (Words) | Admiralty Code (Letters) |
| **High Confidence** | well-corroborated info | high quality information | Reliable | A |
| | reliable sources | multiple sources | Usually Reliable | B |
| | low potential for deception | additional sources only refine analytical judgments | | |
| | assumptions not critical | | | |
| | Strong Logical Inferences | | | |
| **Medium Confidence** | partially corroborated info | credibly sourced | Fairly Reliable | C |
| | good sources | plausible | | |
| | moderate potential for deception | not of sufficient quality or corroboration | | |
| | assumptions potentially critical | additional sources have potential to substantively change analytic judgments | | |
| | mix of strong and weak inferences | | | |
| **Low Confidence** | uncorroborated information | information's credibility or plausibility is uncertain | Not Usually Reliable | D |
| | marginal sources | information too fragmented | Unreliable | E |
| | high potential for deception | information is poorly corroborated | | |
| | key assumptions critical | reliability of sources is questionable | | |
| | weak inferences | analytical judgments should be considered preliminary | | |

## 10.3.   Review of Existing Data Sharing Structures:

The focal node for this review is the SIEM (Security Information and Event Management) list of software products, maintained at *Nigel Willson Cyber IT Security Strategist's online blog (Mar. 14, 2014).*

That source lists the following "frameworks, tools, standards, and working groups:"

- **OpenIOC** – Open Indicators of Compromise framework
- **VERIS** – Vocabulary for Event Recording and Incident Sharing
- **CybOX** – Cyber Observable eXpression
- **IODEF** – Incident Object Description and Exchange Format
- **TAXII** – Trusted Automated eXchange of Indicator Information
- **STIX** – Structured threat Information Expression
- **MILE** – Managed Incident Lightweight Exchange
- **TLP** – Traffic Light Protocol
- **OTX** – Open Threat Exchange
- **CIF** – Collective Intelligence Framework

*Table 10-3* provides a brief description of each effort. They are characterized as dealing with the "semantic" – that is, the meaning of reports about incidents, or "technical" – that is, dealing with the software or hardware aspects of event management. The author is not a technical expert, and did not attempt to annotate the more technical efforts.

Table 10-3. Overview on Information Sharing Efforts, either Semantic (Sem.) or Technical (Tech.).

| | Brief Description | Sem. | Tech. | Annotation |
|---|---|---|---|---|
| OpenIOC | OpenIOC is an extensible XML schema that enables you to describe the technical characteristics that identify a known threat, an attacker's methodology, or other evidence of compromise. | | X | Seems to have only a rudimentary interface |
| VERIS | Vocabulary for Event Recording and Incident Sharing | X | | Built around: incident tracking, victim demographics, incident description, incident details, discovery & response, & impact assessment using an online survey |
| CybOX | Cyber Observable eXpression | | X | Straight up code and mentions that its latest has been integrated into STIX |
| IODEF | Incident Object Description and Exchange Format: I provides an XML representation for conveying incident information across administrative domains between parties that have an operational responsibility of | | X | |

| | | | X | |
|---|---|---|---|---|
| | remediation or a watch-and-warning over a defined constituency. | | | |
| TAXII | Trusted Automated eXchange of Indicator Information | | X | |
| STIX | Structured threat Information Expression | | X | |
| MILE | Managed Incident Lightweight Exchange | | X | There seems to be a standards group of which MILE is a component |
| TLP | Traffic Light Protocol: Possibly created by NISCC (National Infrastructure Security Security Co-ordination Center) | X | | |
| OTX | Open Threat Exchange | X | | This seems to be pushing for a very attractive interface for their tool. |
| CIF | Collective Intelligence Framework | | X | |

We paraphrase briefly here some comments related to these efforts, from expert interviews (E-numbers), about either SIEM or SIEM projects that aim to incorporate confidence measures.

Regarding STIX (Structured Threat Information Expression) and TAXXI (Trusted Automated Exchange of Indicator Information), E1 stated that they are a little light on things like where is the information coming from and how is it assessed. They rely on things like, "I trust this information because it is coming from Fred and Fred is in the circle." E2 mentioned IBM or Nitro SIEM technology (number of events and severity) and that MS-ISAC and FS-ISAC are moving towards STIX and TAXXI -- but for the ISACS it's still a manual product.

From our literature search De Fuentes, *et Al.* (2017) write "In this paper we introduce PRACIS, a scheme for CIS networks that guarantee private data forwarding and aggregation. PRACIS leverages the well-known Structured Threat Information Expression (STIX) standard data format."

E1 also mentioned that GTRI is working to provide a formatting method to handle issues like authenticating that "Person X is a sworn officer in jurisdiction Y" by providing look-up tables and small encapsulated messages[2]. E1 mentioned that SAML has been used in the private sector for over ten years to carry assertions and identity information, stating further that it was invented "before its time." Also, the ORMS project (Open Reputation Management Systems project in OASIS) began more than five years ago. E5 & E6 mentioned work at John Hopkins Advanced Physics Lab on Cybersecurity and E1 suggested looking into AIS (Automatic Indicator Sharing), Provenance, and Assertions.

**Obstacles Regarding Data Sharing Structures:**

NIST in their SP 800-150 Guide to Cyber Threat Information Sharing mention (p.5)

---

[2] Web site: https://trustmark.gtri.gatech.edu/

"Some information sharing challenges apply only to the consumption of threat information:

•**Accessing External Information**. Organizations need the infrastructure to access external sources and incorporate the information retrieved from external sources into local decision-making processes. Information received from external sources has value only to the extent that an organization is equipped to act on the information.

•**Evaluating the Quality of Received Information**. Before acting on threat information, an organization needs to confirm that the information is correct, that the threat is relevant, and that the risks of using or not using the information (i.e., potential impacts of action vs. inaction) are well understood."

Expert Comments:

E2 mentioned that DHS moved to recommend a standard: STIX, to be used in the TAXII environment. NCICC embraced it -- but E2 asserted that the partners "don't really get it." Expert E2 publishes in that format, and gives outside partners a code snippet to extract what they need from a TAXII file, so they can read it and use it.

## 10.4.    Selected Directives Regarding Probabilities

### 10.4.1. Intelligence Community Directive 203 Analytic Standards

(a) For expressions of likelihood or probability, an analytic product must use one of the following sets of terms:

| Almost no chance | Very unlikely | Unlikely | Roughly even chance | Likely | Very likely | Almost certain(ly) |
|---|---|---|---|---|---|---|
| Remote | Highly improbable | Improbable (improbably) | Roughly even odds | Probable (probably) | Highly probable | Nearly certain |
| 01-05% | 05-20% | 20-45% | 45-55% | 55-80% | 80-95% | 95-99% |

Figure 3, ODNI Guidelines on Confidence.

Analysts are strongly encouraged not to mix terms from different rows. Products that do mix terms must include a disclaimer clearly noting the terms indicate the same assessment of probability.

(b) To avoid confusion, products that express an analyst's confidence in an assessment or judgment using a "confidence level" (e.g. "high confidence") must not combine a confidence level and a degree of likelihood, which refers to an event or development, in the same sentence.

### 10.4.2. USCG "What We Mean When We Say" Document

**WHAT WE MEAN WHEN WE SAY: AN EXPLANATION OF ESTIMATIVE LANGUAGE**

We use phrases such as we judge, we assess, and we estimate – and probabilistic terms such as probably and likely – to convey analytical assessments and judgments. Such statements are not facts, proof, or knowledge. These assessments and judgments generally are based on collected information, which is often incomplete or fragmentary.

Some assessments are built on previous judgments. In all cases, assessments and judgments are not intended to imply that we have "proof" that shows something to be a factor that definitively links two items or issues.

In addition to conveying judgments rather than certainty, our estimative language also often conveys 1) our assessed likelihood or probability of an event; and 2) the level of confidence we ascribe to the judgment.

**ESTIMATES OF LIKELIHOOD**: Because analytical judgments are not certain, we use probabilistic language to reflect estimates of the likelihood of developments or events. Terms such as "probably," "likely," "very likely," or "almost certainly" indicate a greater than even chance. The terms "unlikely" and "remote" indicate a less than even chance that an event will occur; they do not imply that an event will not occur. Terms such as "might" or "may" reflect situations in which we are unable to assess likelihood, generally because relevant information is unavailable, sketchy, or fragmented. Phrases such as "we cannot dismiss," "we cannot rule out," or "we cannot discount" reflect an unlikely, improbable, or remote event whose consequences are such that it warrants mentioning. The following chart provides a rough idea of the relationship of some of these terms to each other.

**CONFIDENCE LEVEL STATEMENTS**: Our assessments and estimates are supported by information that varies in scope, quality, and sourcing. Consequently, we ascribe high, moderate, or low levels of confidence to our assessments, as follows:

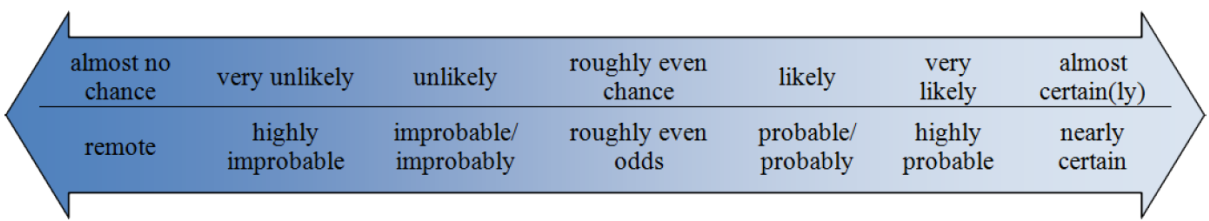

Figure 4. The ordering of USCG Estimative Terminology

| | |
|---|---|
| *High Confidence* | • Well-corroborated information from reliable sources<br>• Low potential for deception exists<br>• Assumptions are not critical to analysis<br>• Reasoning dominated by strong logical inferences developed through multiple analytic techniques |
| *Moderate Confidence* | • Partially corroborated information from good sources<br>• Moderate potential for deception exists<br>• Assumptions are potentially critical to analysis<br>• Reasoning with a mixture of strong and weak inferences developed through simple analytic techniques |
| *Low Confidence* | • Minimally or uncorroborated information from good or marginal sources<br>• High potential for deception exists<br>• Key assumptions are critical to analysis<br>• Reasoning dominated by weak inferences developed through few analytic techniques |

Figure 5. Details of USCG Estimative Terminology

### 10.4.3. Input from the FBI:

**(U) Expressions of Likelihood (or Probability)**

(U) Phrases such as "the FBI judges" and "the FBI assesses," and terms such as "likely" and "probably" convey analytical judgments and assessments. The chart approximates how expressions of likelihood and probability correlate with percentages of chance.

UNCLASSIFIED

| *Terms of Likelihood* | Almost No Chance | Very Unlikely | Unlikely | Roughly Even Chance | Likely | Very Likely | Almost Certain(ly) |
|---|---|---|---|---|---|---|---|
| *Terms of Probability* | Remote | Highly Improbable | Improbable (Improbably) | Roughly Even Odds | Probable (Probably) | Highly Probable | Nearly Certain |
| | 1-5% | 5-20% | | | | | |

Figure 6. Confidence Scale Used by the FBI

(U) Unless otherwise stated, the FBI does not derive judgments via statistical analysis.

**(U) Confidence in Sources Supporting Assessments and Judgments**

(U) Confidence levels reflect the quality and quantity of the source information supporting judgment. Consequently, the FBI ascribes high, medium, or low levels of confidence to assessments, as follows:

(U) **High confidence** generally indicates the FBI's judgments are based on high quality information, from multiple sources. High confidence in a judgment does not imply the assessment is a fact or a certainty; such judgments might be wrong. While additional reporting and information sources may change analytical judgments, such changes are most likely to be refinements and not substantial in nature.

(U) **Medium confidence** generally means the information is credibly sourced and plausible but not of sufficient quality or corroborated sufficiently to warrant a higher level of confidence. Additional reporting or information sources have the potential to increase the FBI's confidence levels or substantively change analytical judgments.

(U) **Low confidence** generally means the information's credibility or plausibility is uncertain, the information is too fragmented or poorly corroborated to make solid analytic inferences, or the reliability of the sources is questionable. Absent additional reporting or information sources, analytical judgments should be considered preliminary in nature.

### 10.4.4. Admiralty Code

The source here is US Army documents that have been posted on the Internet by the Federation of Atomic Scientists, at: https://fas.org/irp/doddir/army/fm2-22-3.pdf . They are shown in Figure 7 and Figure 8

**Source and Information Reliability Matrix**

**SOURCE RELIABILITY**

B-1. Reliability ratings range from "Reliable" (A) to "Unreliable" (E) as shown in Table B-1. In every instance the rating is based on previous reporting from that source. If there has been no previous reporting, the source must be rated as "F". [NOTE: An "F" rating does not necessarily mean that the source cannot be trusted, but that there is no reporting history and therefore no basis for making a determination.]

Table B-1. Evaluation of Source Reliability.

| A | Reliable | No doubt of authenticity, trustworthiness, or competency; has a history of complete reliability |
| B | Usually Reliable | Minor doubt about authenticity, trustworthiness, or competency; has a history of valid information most of the time |
| C | Fairly Reliable | Doubt of authenticity, trustworthiness, or competency but has provided valid information in the past |
| D | Not Usually Reliable | Significant doubt about authenticity, trustworthiness, or competency but has provided valid information in the past |
| E | Unreliable | Lacking in authenticity, trustworthiness, and competency; history of invalid information |
| F | Cannot Be Judged | No basis exists for evaluating the reliability of the source |

**INFORMATION CONTENT**

B-2. The highest degree of confidence in reported information is given to that which has been confirmed by outside sources, "1". Table B-2 shows evaluation of information content. The degree of confidence decreases if the information is not confirmed, and/or does not seem to make sense. The lowest evaluated rating of "5" means that the information is considered to be false. [NOTE: A rating of "6" does not necessarily mean false information, but is generally used to indicate that no determination can be made since the information is completely new.]

Figure 7. The Admiralty Code for Confidence

_________________________________________________

**Table B-2. Evaluation of Information Content.**

| 1 | Confirmed | Confirmed by other independent sources; logical in itself; Consistent with other information on the subject |
|---|-----------|---------------------------------------------------------------------------------------------------------------|
| 2 | Probably True | Not confirmed; logical in itself; consistent with other information on the subject |
| 3 | Possibly True | Not confirmed; reasonably logical in itself; agrees with some other information on the subject |
| 4 | Doubtfully True | Not confirmed; possible but not logical; no other information on the subject |
| 5 | Improbable | Not confirmed; not logical in itself; contradicted by other information on the subject |
| 6 | Cannot Be Judged | No basis exists for evaluating the validity of the information |

Figure 8. The Admiralty Numerical Scoring Rules

## 10.5. Summary of Literature Examined

The following materials were selected as likely to be valuable for further research into the issues of asserting and/or acquiring confidence.

This page intentionally blank

## 11. Appendix B: The Modified Online Delphi Process

Prepared by Katie L. McKeon

### 11.1. Summary of the Process

The Modified Online Delphi Process includes four rounds and allows any number of expert panelists to participate remotely via Google Sheets using any Internet browser. During the session, the panelists[3] also need to join a conference call. The facilitator, through the conference call, guides experts through each round and designates portions where the experts are allowed to discuss results with each other.

There are two types of rounds: evaluation and suggestion. Rounds one, three, and four are evaluation rounds. Round one asks participants to evaluate a list of practices given a set list of criteria. The suggestion round (round two) asks participants to suggest practices that they believe were missing from the list in round one. Round three asks for evaluation of potential policies. Finally in round four, participants evaluate the list of practices that they generated in round two.

For each round, the participants will find a worksheet corresponding to that round within their Google spreadsheets. Note that round four is optional, so participants will not see that worksheet until the group agrees to proceed to round four. In the evaluation rounds, participants select all the parameters which apply to each practice (for instance, 'is it feasible?') Once all the participants have finished, the facilitator reveals the total votes and guides a discussion on the results. The participants are allowed and encouraged to rethink their responses and change it if necessary. In the suggestion round, participants will enter all of the practices that they would like to recommend and once everyone has finished the list of all suggestions will be displayed. At this time, the facilitator will guide discussion in order to organize the list, eliminate redundancies, and possibly add more suggestions that occur during the discussion.

The use of Google Sheets and Google Scripts, which power many of the features in the spreadsheets, allows the facilitator to easily tally all of the votes in the evaluation rounds, gather suggestions in round two, and quickly generate the worksheets required for round four. The next section describes the three different types of spreadsheets which are involved in the Modified Online Delphi Process. The final section describes a tool which can automatically design and link these spreadsheets.

### 11.2. Spreadsheets

There are three types of spreadsheets: those for participants, those for the facilitator, and those for observers. Participants use their spreadsheets to enter evaluations and make suggestions. The facilitator uses a '*mastersheet*' to supervise them as they participate and to export any relevant data to all of the participants sheets. More specifically, the *mastersheet* can:

- aggregate all the evaluations and send the totals to the experts
- retrieve suggestions from experts in round two, compile the list, and send it to the experts
- generate the worksheet for round four for each expert as well as the facilitator

---

[3] In the remainder of this note the panelists are referred to as "participants," as is customary in the social sciences.

Finally, the *observer* sheet allows any team member to observe the session as it takes place. The observer sheet streams a copy of the mastersheet. This allows team members to participate in the session remotely and to see the participant input without altering anything in the mastersheet.

The following sections provide a more detailed description of the worksheets within the participants spreadsheet and the mastersheet.

### 11.2.1. Participant Sheet

Before the session, each of the participants is sent a link to a personal spreadsheet which will allow them to participate in the session. The participants are given identifiers such as 'Expert A', 'Expert B', etc. and the links are sent individually in order to preserve anonymity.

*Round One*

In this round, participants evaluate a list of practices according to a fixed set of criteria. The participant selects any parameter which applies by placing an X in the corresponding cell. In the figure below, the expert is asked to consider the following criteria:

- 
- Have Tried it (H)    Do you have personal experience with this practice?
- Feasible (F)        Is the practice realistic and possible to implement?
- Sustainable (S)    Can the practice be sustained indefinitely?
- Important (I)
- Not Important (N)

The participants may enter any comments in the last column. Once all the participants have finished entering their evaluations, the facilitator tallies the results, which are displayed in the lower right corner of the expert's worksheet (the grey cells in the Figure 11-1.)

## Layout of Round One Worksheet

**Round Two**

In Round Two, participants are asked to suggest additional practices. In Column A, the participant types any practices that they did not see in Round One. Once all participants are finished making suggestions, the facilitator uses the mastersheet to compile the list. The full list of suggestions will be displayed in Column B of expert's Round Two worksheet.

## Layout of Round Two Worksheet

*Round Three*

In Round Three, participants are asked to evaluate certain proposed policies. The participant's worksheet is laid out similar to Round One. However, the expert is asked to evaluate policies according to a slightly revised set of parameters:

- Have Tried it (H)
- Feasible (F)
- Important (I)
- Not Important (N)

Once all the participants are finished evaluating the policies, the facilitator tallies results so that the group may discuss them.

*Round Four*

In Round Four, participants evaluate the list of suggestions that they created in Round Two. The criteria that they consider are, by default, the same as in Round One.

## 11.2.2. Mastersheet

The mastersheet contains functions that affect all the other sheets. As such, it is the most complex spreadsheet. The following sections describe not only the layout of the mastersheet, but also the functions, which are accessible through the menu bar of the spreadsheet.

*Round One*

The mastersheet allows the facilitator to observe work of the participants as they enter their evaluations. The leftmost columns contain the practices, the middle columns automatically update to show what the specific participants have entered, and the rightmost columns contain space for tallied votes. The facilitator has three menu options in this round: **Tally Results**, **Hide Individual Expert[4] Columns**, and **Show Individual Expert Columns**.

The facilitator selects **Tally Results** once all the participants have finished evaluating the practices. This function totals the criteria selected for each practice among all the participants and then exports the totals to the participant sheets. Note that comments are not sent out to the participants. Rather, the comments (along with the totals) are exported to a log document to be used by researchers.

The functions **Hide** and **Show** Individual Expert Columns allow the facilitator to condense the view of the worksheet. If there are many participants in the session, the facilitator would have to scroll left and right on the mastersheet to move between the tallied results and the practices that they correspond to. So **Hide Individual Expert Columns** collapses all of the middle columns corresponding to Expert A, Expert B, etc. Conversely, **Show Individual Expert Columns** returns the sheet to its original form.

**Layout of Round One Worksheet**

---

[4] In the spreadsheet itself the term "Expert" is used as the participants or panelists in a Delphi process are customarily subject matter experts.

*Round Two*

The master worksheet for this round compiles suggestions for practices from the participants. The mastersheet will *not* show suggestions as the participants are entering them. So, it is necessary for the facilitator to verbally check whether the participants have finished entering their inputs. There are two menu items for this round: **Gather Suggestions** and **Update Suggestion List.**

Gather Suggestions retrieves all the inputs from the participants. Then it alphabetizes the list[5] and displays it in Column A of the mastersheet. It also exports the same list to Column B of each participant sheet as well as to the log document.

As discussion progresses in Round Two, the facilitator may choose to add, delete, or combine certain items. After any such revisions are made in column A of this worksheet, the facilitator selects the **Update Suggestion List** function to export the new list to all the participants.

**Layout of Round Two Worksheet**

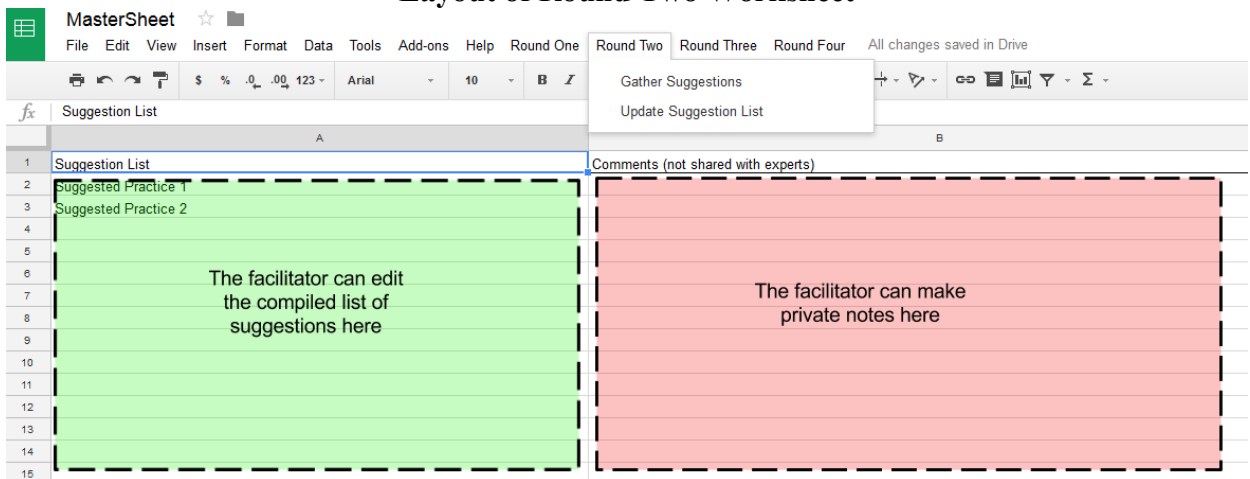

*Round Three*

The Round Three worksheet is nearly identical to Round One in layout and functionality. Recall that, in this study, participants evaluate policies in this round. So, the evaluation criteria change slightly.

*Round Four*

The Round Four worksheet, collects all the suggestions that were made in Round Two (the final version of Column A of the Round Two worksheet tab in the mastersheet) and assembles another round of evaluation. The style of this round is again identical to Round One. Only the list of practices has changed. There are four menu items for Round Four: **Initiate Round Four**, **Tally Results**, **Hide Individual Expert Columns**, and **Show Individual Expert Columns**.

If the Facilitator elects to Initiate Round Four, then a new worksheet tab is created on each of the participant sheets as well as the mastersheet. The **Initiate Round Four** function has already been executed in Figure 11-5.

---

[5] This step facilitates discussion and will generally separate suggestions from a single participant.

The other three menu items for Round Four (**Tally Results**, **Hide Individual Expert Columns**, and **Show Individual Expert Columns**) serve the same function as those for Round One.

## Layout of Round Four Worksheet

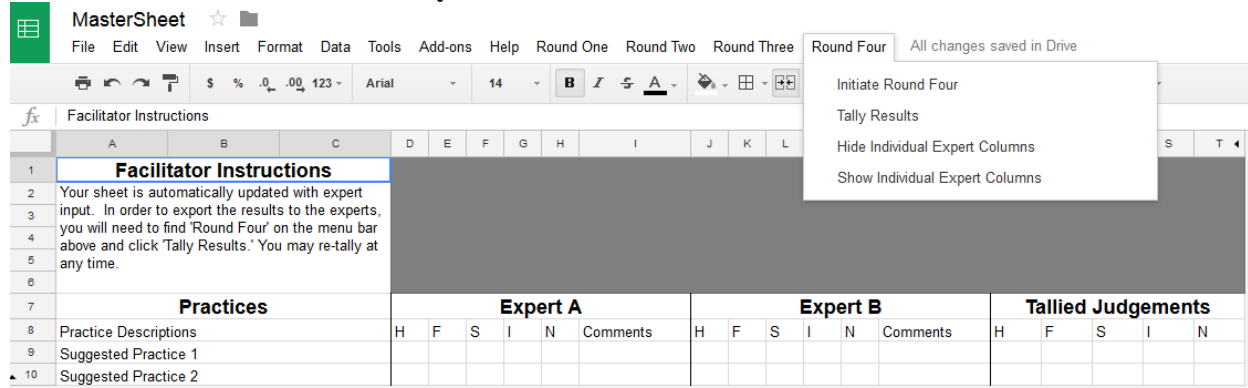

## 11.3. Modified Online Delphi Generator

This section contains a description of the tool that creates all of the spreadsheets for the Modified Online Delphi Session. The tool creates any number of participant sheets along with a corresponding mastersheet. The layout and functionality of the resulting spreadsheets are identical to the sheets described in the previous section. The 'Delphi Generator' tool accomplishes this task in seconds. Note that the Delphi Generator is also a Google Sheet, using Google Scripts to design and create the expert sheets and mastersheet. Figure 11-7 at the end of this section provides the layout of the Delphi Generator.

The researcher who wishes to initiate a session first selects the number of participants to participate, as well as the number of observers. Next, the user enters information about Round One. First, the user provides the criteria which participants should consider. Second, provide a list of practices to evaluate. There is no information required for Round Two. For Round Three, the researcher should enter a list of policies and then the parameters (responses) that are allowed for that round. There is no required information for Round Four. Note that the generator assumes that the same criteria from Round One still apply for Round Four.

Having entered all the required information, the user selects '**Delphi: Generate Sheets**' from the menu bar.

## Menu Bar for the Tool

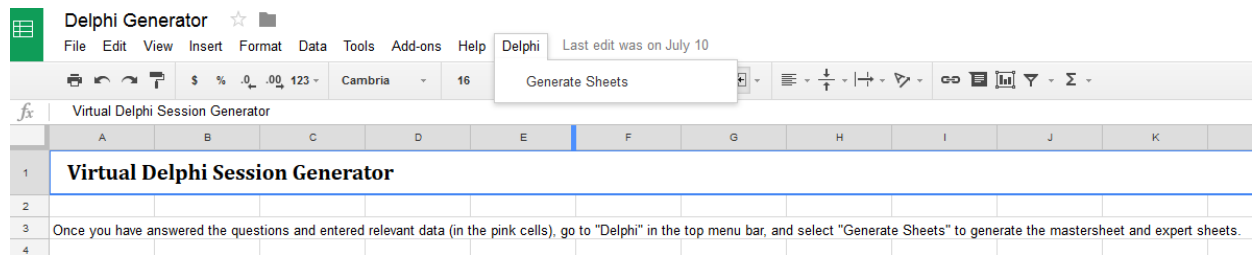

All of the required sheets (and the log document) will be stored in the user's Google Drive. Once created, the sheets can be moved to a different folder (within the same user's Google Drive) and even renamed without modifying the links between the master sheet and the expert sheets. However, the contents and layout of the worksheets should not be altered, as certain functions read from specific ranges in specific sheets.

After the sheets have been created, the facilitator, or whoever generates the sheets, must enable sharing of the participant sheets, so that participants may access their individual spreadsheets without being required to log into a Google account. This can be done by selecting each participant sheet, and finding 'File: Share' in the menu bar. The facilitator should ensure that "anyone with a link" can edit the participant sheet, and record the link so that it can be sent to a participant. The sharing process applies also to all the observer sheets.

The mastersheet requires that the facilitator be logged in to the same Google account where is stored. After the first function is selected, the facilitator will have to grant permission (from this account) to run the function. This is because the participant sheets in the same user's directory are being accessed and modified during the Delphi process.

# Example Data for the Delphi Generator

## Virtual Delphi Session Generator

Once you have answered the questions and entered relevant data (in the pink cells), go to "Delphi" in the top menu bar, and select "Generate Sheets" to generate the mastersheet and expert sheets.

## Experts & Observers

| | Number of Experts | 2 | | | Number of Observers | 0 |
|---|---|---|---|---|---|---|

## Evaluation Parameters

| | Instructions | In each column, you will enter a parameter, a single character to describe it, and (optional) a longer explanation. | | | |
|---|---|---|---|---|---|
| | Example | Parameter | Feasible | | |
| | | Character | F | | |
| | | Description | Is it possible to accomplish it? | | |

| Parameter | Have Tried it | Feasible | Sustainable | Important | Not Important |
|---|---|---|---|---|---|
| Character | H | F | S | I | N |
| Description | Do you have pers | Is the practice re | Can the practice be sustained indefinitely? | | |

## Group & Practice Descriptions

| | Instructions | In each column, you will enter a practice, a group or category which it fits into, and (optional) a longer explanation. | | | | | | | | | |
|---|---|---|---|---|---|---|---|---|---|---|---|
| | Example | Practice | Report MAC | | | | | | | | |
| | | Group | Provenance | | | | | | | | |
| | | Description | Reporting the mac address that a message originates from | | | | | | | | |

| Group | Asserting | Asserting | Asserting | Asserting | Asserting | Asserting | Asserting | Asserting | Acquiring | Acquiring | Acquiring |
|---|---|---|---|---|---|---|---|---|---|---|---|
| Practice | ASSTN_1: Send | ASSTN_2: Hand | ASSTN_3: Reim | ASSTN_4: Take | ASSTN_5: Chec | ASSTN_6: Chec | ASSTN_7: Chec | ASSTN_8: Chec | ACQRG_1: Act | ACQRG_2: ONL | ACQRG |

## Policy Descriptions

| | Instructions | In each column, you will enter a practice, a group or category which it fits into, and (optional) a longer explanation. | | | |
|---|---|---|---|---|---|
| | Example | Practice | Report MAC | | |
| | | Group | Provenance | | |
| | | Description | Reporting the mac address that a message originates from | | |

| Group | Policy Level Prac | Policy Level Prac | Policy Level Prac | Policy Level Prac | Policy Level Practices |
|---|---|---|---|---|---|
| Practice | POL_1 DHS sho | POL_2 Some ch | POL_3 The comi | POL_4 Report ex | POL_5 STIX should have a way of embedding supporting evidence in the indicator |

## Policy Evaluation Parameters

| | Instructions | In each column, you will enter a parameter, a single character to describe it, and (optional) a longer explanation. | | | |
|---|---|---|---|---|---|
| | Example | Parameter | Feasible | | |
| | | Character | F | | |
| | | Description | Is it possible to accomplish it? | | |

| Parameter | Have Tried it | Feasible | Important | Not Important |
|---|---|---|---|---|
| Character | H | F | I | N |
| Description | Do you have pers | Is the practice realistic and possible to implement | | |

# 12. Appendix C: A Theoretical Discussion of the Uses of Confidence

Prepared by Paul B. Kantor

## 12.1. Introduction

Any method for communicating confidence must include at least two ingredients: rules for assigning a particular level of confidence, to the information being transmitted; and rules for using the message when it is received. If messages always refer to unique and unrelated events, then a single set of rules suffices. However, modern cyber-security efforts emphasize the integration of multiple messages about threats, as a key step in identifying the scope and threat of a particular exploit, and its possible relation to larger campaigns.

Both the literature and subject matter experts suggest that there are at least two dimensions to the concept of confidence. One dimensions relates to "the probability that the information is a correct description of some specific event." The earliest references to this dimension seems to lie in the Intelligence Community (IC) and were initiated in a famous (although classified) essay (Kent, 1963). It is reported (at the cited site) that the ten-fold classification that he proposed was not widely adopted. Agencies sharing cyber-security information are working to draw up usable and clear scales. In the US a policy developed at the ODNI appears to be widely disseminated and in use (Office of the Director of National Intelligence, 2015). This policy defines six ranges, which are symmetrical, and assigns a natural language term to each range. While this appears to be less precise than Kent's original proposal, it may better reflect the intrinsic difficulty of trying to be precise about uncertainty.

One might imagine that using a single term[6] that includes probabilities ranging from five percent (a little better than the chance of picking a red ace from a complete deck of cards) to almost 20% (more than the chance of rolling a six with a single die) would make experts uncomfortable. In fact, however, no expert, and none of the literature that was consulted indicated any such discomfort.

An additional important component of confidence is captured in the scale adopted by the financial services information sharing and analysis center (FS-ISAC). This scale, adopted from the UK Admiralty, reports on two dimensions that are known to be important. One is the estimate of certainty itself, which corresponds, conceptually, to the ODNI Directive (see Section 10.4.1. The other dimension estimates the reliability of the source from which the report has been received. This is clearly important when several related reports are to be combined. Intuitively, multiple reports of very similar threats raise the significance of the threat, and multiple reports about the confidence, or probability that the threat is real should sharpen the estimate of confidence.

We have not been able to determine the order in which these two concepts are applied, and it seems likely that there may be divergences in practice. One possibility is that the reliability of the source is "baked into" the confidence estimate. In this case, no report originating with an unreliable source could have high confidence. The other possibility is that the report itself relays an estimate of confidence, which is then modulated by recipient's assessment of the reliability of the source. The second alternative seems likely, in cases where reports come from human actors, who are known to have complex motives. However, as the sharing of information about cyber-threats becomes more nearly automatic, a report may originate with an algorithm, which not only reports a confidence level, but also knows something about its own "track record." In effect, such an algorithm might say "I have 80% confidence in this report, but I have been wrong about 15% of the time, in the past."

---

[6] In the ODNI system, the term: "very unlikely" includes probabilities from 5% to 20%.

While we have not found any literature on the problem of automatically combining estimates of confidence and reliability, we are confident that such processes will eventually be required. The remainder of this discussion examines some possible ways of dealing with the two-dimensional nature of confidence.

## 12.2.    Combining Reports of Confidence with Assessments of Reliability

To begin, we note that both scales assess a quantity that is a probability: roughly speaking "the probability that the report is true." Although each report is a single instance, the concepts of probability can be applied, much as they are applied to assessing the accuracy of a weather service when it predicts "a 70% chance of rain." To avoid excessive mathematics, we concentrate on specific assessments, which are quintiles for the probability.

We can label the quintiles as: [0-20%] =Q1, Q2, …, Q5=[81%-100%]. Now, suppose we have received two reports: Q1 (with reliability = highest) and Q3 (also with reliability = highest). We need some heuristic rule for deciding how much confidence we should have, in this situation.

Let us agree that highest reliability also means "in the top 20% of reliability[7]." We may interpret that by saying that there is an 80% chance that the confidence should be as it is reported. In other words, there is a 20% chance that it is something else. While that other confidence might be anything at all, it seems again sensible to propose that it is "nearby." So a highest reliability report of Q3 might be conceptualized as shown in *Table 12-1*

Table 12-1. A high reliability report of confidence between 60% and 80%

|          | Q1 | Q2  | Q3  | Q4  | Q5 |
|----------|----|-----|-----|-----|----|
| Report 1 |    | 10% | 80% | 10% |    |

Similarly, the report of "Q1 with highest reliability" adds another row.

Table 12-2. A high reliability report of confidence between 0% and 20%.

|          | Q1  | Q2  | Q3  | Q4  | Q5 |
|----------|-----|-----|-----|-----|----|
| Report 1 |     | 10% | 80% | 10% |    |
| Report 2 | 80% | 20% |     |     |    |

---

[7] The Admiralty does not provide numeric definitions, but any algorithm for combining reports will require this kind of numerical translation of the terms used.

In *Table 12-2*, we have placed all of the "leftover probability in the nearest confidence estimate. Another approach might spread it out over two alternatives. There is no solid mathematical reason to prefer one over the other. There is some literature suggesting that when humans assign very high or very low confidence they are more likely to be wrong. This observation might lead to a preference for spreading extreme estimates more broadly, as shown in this *Table 12-3*.

Table 12-3. An alternative way of dealing with extreme assessments of confidence

|  | Q1 | Q2 | Q3 | Q4 | Q5 |
|---|---|---|---|---|---|
| Report 1 |  | 10% | 80% | 10% |  |
| Report 2 | 80% | 20% |  |  |  |
| Alt: Report 2 | 80% | 10% | 10% |  |  |

For this discussion we will stick with the first alternative. This provides a basis for looking at several ways to combine "independent pieces of evidence."

The first approach to combination asks for the probability of either of two events. It might be translated as "the probability that the confidence is Q3 because of Report 1 or that it is Q3, because of Report 2."

As shown in Case A, this rule behaves quite sensibly for the conflicting reports of this example[8]. Each of the reported possibilities keeps its strength, but the compromise, which has spillover from both of them, also picks up strength, being over 1/3 as "strong" as either of the asserted confidence levels[9].

Table 12-4. Comparing methods of combination

| Logical OR | | | | | |
|---|---|---|---|---|---|
| CASE A | Q1 | Q2 | Q3 | Q4 | Q5 |
| Report 1 |  | 10% | 80% | 10% |  |
| Report 2 | 80% | 20% |  |  |  |
| Combined | 80% | 28% | 80% | 10% | 0% |
|  |  |  |  |  |  |
| Logical OR | | | | | |
| CASE B | Q1 | Q2 | Q3 | Q4 | Q5 |
| Report 1 |  | 10% | 80% | 10% |  |
| Report 3 |  | 10% | 80% | 10% |  |
| Combined | 0% | 19% | 96% | 19% | 0% |
| Odds Multiplication | | | | | |
| CASE C | Q1 | Q2 | Q3 | Q4 | Q5 |
| Report 1 |  | 10% | 80% | 10% |  |
| Report 3 |  | 10% | 80% | 10% |  |

---

[8] The specific formula for combining two probabilities, *p* and *p'* is : $p_{combined}=1-(1-p)(1-p')$
[9] The results are actually numbers, rather than percentages. However, all that matters is their relative sizes.

| | | | | | |
|---|---|---|---|---|---|
| Odds 1 | 0.000 | 0.111 | 4.000 | 0.111 | 0.000 |
| Odds 2 | 0.000 | 0.111 | 4.000 | 0.111 | 0.000 |
| Product | 0.000 | 0.012 | 16.000 | 0.012 | 0.000 |
| Probability | 0% | 1% | 94% | 1% | 0% |
| Odds Multiplication | | | | | |
| CASE D | Q1 | Q2 | Q3 | Q4 | Q5 |
| Report 1 | | 10% | 80% | 10% | |
| Report 2 | 80% | 20% | | | |
| Odds 1 | 0.000 | 0.111 | 4.000 | 0.111 | 0.000 |
| Odds 2 | 4.000 | 0.250 | 0.000 | 0.000 | 0.000 |
| Product | 0.000 | 0.028 | 0.000 | 0.000 | 0.000 |
| Probability | 0% | 3% | 0% | 0% | 0% |

However, when we apply the same rule to a pair of reports that really agree with each other (Case B), the results are not quite so satisfying. While the strength of the shared assertion grows, as we would expect, the two neighboring possibilities also gain strength, so that they seem relatively stronger than they did in the individual reports. This follows from the mathematics, but it does not seem intuitively correct.

We can contrast this with a method called "combination of odds ratios" (also called "Bayesian updating" (Pearl, 2014)). The mathematics for that is somewhat more complicated. Each probability is converted to an odds ratio (for example 75% probability becomes 75%:25% = 3.0). These ratios, as asserted by the several reports, are multiplied together, and finally the answer is converted back to a probability. As shown in Case C, this method behaves very well in the case of two agreeing reports. The common assessment winds up with a weight of 94%, while the neighbors, which were only there because the reports are not 100% reliable, are suppressed.

However, when we come back to the case of two reports that don't agree (Case D), odds multiplication completely annihilates both of the reported values, a result which many find to be intuitively unsatisfying. The problem that we have sketched here is well known in probability and decision theory, and is not presently resolved. A seminal discussion of these problems is found in (Good, 1983). Some experts believe that an approach called Dempster-Shafer Theory will resolve the problem (Sentz and Ferson, 2002). However, there are no successful implementations of this theory, and the approach tends to require enormous sets of possibilities, as more reports are received.

## 12.3. Independence, Provenance and the Echo Chamber

All of the discussion so far has treated the two reports that are being combined as being independent pieces of information. If this condition is not met, it is very hard to devise a suitable formula for combining the reports. The danger is that the reports will generate an echo chamber, in which the same fundamental piece of information is repeated many times. Rather than seeking a solution in very complex mathematics, cyber-security information sharing should seek to assess and maintain independence of the reports.

This leads directly to the issue of the "provenance of a report." A report, when it is received, appears to come directly from its author, which may be a human actor, or a system of some kind, or both. But that metadata rarely reveals the source(s) of other information that has fed into the report, and influenced its conclusions. This is the problem known generally as "provenance tracking" (Abiteboul et al., 2000; Buneman et al., 2001) and it has been a subject of interest to the IC for a number of years (IARPA, 2013). Based on the fact that this problem remains open, the cybersecurity community might choose to look for a "practical fix." As an example, one key step would be to make sure that every report is clearly labelled as either "a new unique event" or "an additional report about an already known event."

## 12.4. Feedback and Automatic Reliability

In the hypothetical example above we described an algorithm that "knows its own reliability." For this to happen, the recipients of reports, from that algorithm, must have a way of reporting back "upstream" when the conclusions that have been reached, based on those reports, turn out to be incorrect. This is related to the problem of provenance, since the path back must be accessible, even if it is not reported in full, to downstream recipients. However, knowing that path is not sufficient. If several pieces of information have been merged to reach an incorrect conclusion, the source of the problem might be the merging process, or any particular subset of the information received. This is, so far as we can determine, a problem that has not been studied, and one that will become important as the infrastructure for sharing information about cyber security is built out.

## 12.5. Conclusions

In sum, whatever principles are developed to guide the reporting of confidence in cyber-security incidents, they should definitely recognize that the reports will eventually be used in automated compilation, either for dissemination or action. It is strongly suggested that those experts who will design the compilation or aggregation be included in discussions about defining the use and meaning of confidence reports. It is also suggested that every report indicate clearly whether it is an initiating report of an incident, or a descendant or secondary report about some incident already reported. Finally, it is suggested that the problem of reporting and using provenance be added to the technical research agenda of the ISE program.

The following materials were distributed to the participants or used by the Facilitator during the Modified Online Delphi Panel process.

## 13.1.    Participant Guide

This guide is designed to illustrate the use of Google Sheets in our facilitated discussion.

You will be given a link to your spreadsheet.  In order for voting to remain anonymous, you should not share the link you've been given or your identifier ('Expert A', 'Expert B', etc.)

As you explore your sheet, notice the two tabs in the bottom left corner entitled 'Round One' and 'Round Two.'  In Round One, we will ask you to evaluate a set of practices and in Round Two you will get the opportunity to share practices of your own.

### 13.1.1. Round One Worksheet

In Columns C through H of your Round One sheet, we'd like you to evaluate each practice with respect to the following criteria:

- Have Tried it (H)      Do you have personal experience with this practice?
- Feasible (F)            Is the practice realistic and possible to implement?
- Sustainable (S)         Can the practice be sustained indefinitely?
- Important (I)
- Not Important (N)

Select all that apply by putting an 'X' in the corresponding cells.  You may also leave comments about the practice and your evaluation which will be shared privately with us.

Once all participants are finished voting, we will view and discuss the results.  After discussion, you will be given the opportunity to change your vote.  You may do so by directly altering the row containing your initial evaluation.  The tallied judgements will not immediately reflect your changed vote.

### Round One Worksheet Layout

### 13.1.2. Round Two Worksheet

In Column A, enter any additional suggested practices that you can think of that we may have missed in the original list. Once all participants are finished making suggestions, these will be shared with the whole group, though they will not be attributed to the individual participants.

Once the suggestions are compiled, the complete list of suggestions will be displayed in Column B. What are your reactions? The goal of discussion in this part is to logically organize and clearly formulate each suggestion. This includes categorizing the practices into groups, and combining similar entries.

**Round Two Worksheet Layout**

### 13.1.3. Round Three Worksheet

This round is nearly identical to Round One. In Columns C through G of your Round Three sheet, we'd like you to evaluate policies with respect to the following criteria:

- Have Tried it (H)     Do you have personal experience with this practice?
- Feasible (F)          Is the practice realistic and possible to implement?
- Important (I)
- Not Important (N)

Select all that apply by putting an 'X' in the corresponding cells. You may also leave comments about the practice and your evaluation which will be shared privately with us.

As in Round One, once all participants are finished voting, we will view and discuss the results. After discussion, you will be given the opportunity to change your vote. You may do so by directly altering the row containing your initial evaluation.

## 13.2. Facilitator Guide

### 13.2.1. 'FAQs'

Q: I can't access my sheet
A: try a different browser

Q: the tallied results didn't show up, they're still grey
A: refresh the browser

Q: What do the acronyms stand for
A: Some ACRONYMS:

VTAPI refers to the Virus Total community: https://api.vtapi.net/hu/documentation/public-api/ a site for sharing phishing or other bad URLs

STIX™, the Structured Threat Information eXpression
TAXII™, the Trusted Automated eXchange of Indicator Information
CybOX™, the Cyber Observable eXpression
See: https://www.us-cert.gov/Information-Sharing-Specifications-Cybersecurity

Q: what if I put other letters in the cells
A: we only count little 'x' and big 'X'

Q: when I scroll down, the criteria heading disappear
A: unfortunately, you'll just have to remember that column C is 'H', column D is 'F', column E is 'S', column F is 'I', and column G is 'S'

### 13.2.2. Session Walkthrough

This guide contains an example walkthrough of a Delphi elicitation session for a facilitator. There are two main roles: a facilitator who will read through the script and handle all communications and a typist who will manage the worksheets. Throughout the rest of this walkthrough, tasks for the facilitator will be highlighted in red and tasks for the typist will be highlighted in blue.

<div align="center">Before the Session</div>

***DO***: Send a link to a different expert sheet for each participant.

<div align="center">Introductory Remarks for Round One</div>

***SAY***: *Welcome to this group discussion, and thank you for your time today. Our discussion today will be using something called a "modified online Delphi process." The name comes from the Oracle at Delphi, and you experts will be our oracles for today's mission. We are working, as we mentioned in the invitation, to identify the most important practices that would help the entire community to communicate the confidence about a threat or attack, and to reason about that confidence information if it comes from multiple sources.*

*SAY*: *If anyone has any technology or process questions, please don't hesitate to ask at any time.*

*SAY*: *Please make sure you are viewing the Round One worksheet of your Expert page. All participants have the same list of groups and practices for review. In column C through F we'd like you to choose from a list of parameters, which ones you feel apply to each practice. The criteria you'll choose from is: "H" for have tried it, "I" for important, "N" for not important, "F" for feasible, and "S" for sustainable. For each parameter which applies, enter an 'X' in that cell.*

## Round One Expert Input

**DO:** Make sure you are starting to see responses appear in your *Round One* tab. Once all participants are finished entering their judgements, it is time to discuss the (anonymized) results with them.

*SAY*: *Now we are going to view the tallied results for each of these practices. I will copy the group of results from my sheet so that they will appear in each of your sheets, under "Tallied Judgements."*

**DO:** Find and click 'Round One: Tally Results' in the menu bar, as shown below. The tallied judgements will be added to the rightmost columns of your Round One sheet. If you do not wish to scroll right and left between the list of practices and the tallied judgements, you may hide the individual expert columns by going to 'Round One' and selecting 'Hide Individual Expert Columns'.

## Round One Discussion

*SAY*: *You should now all be able to see the tallied results in your Round One sheet.*

*SAY*: *Let's see whether any consensus or lack of agreement jumps out at us here. We will go through the list and discuss the results for each practice separately, after which you are allowed and encouraged to rethink your vote.*

**DO:** After experts have discussed a practice, select 'Round One: Tally Results' from the menu bar in order to update the tally and announce to the experts that you've updated it. Only update the tally once for each practice.

## Introductory Remarks for Round Two

*SAY*: *Now we would like to ask you to suggest any additional practices we may have missed. Please note these will be shared with the whole group, though they will not be attributed to the individual experts.*

*SAY*: *Please select the second worksheet tab, Round Two. In the rows of Column A, you may enter any number of suggested practices. You do not need to provide a group that the practice falls in, we'll get to that later.*

<u>Round Two Discussion</u>

***DO:*** Once all participants are finished entering suggestions, select 'Round Two: Gather Suggestions' in the menu bar (as shown below.)  The suggestions will be gathered from all the experts, alphabetized, and displayed in your Column A.  This list will be automatically shared with the experts.

*SAY: Notice that in your Round Two worksheet, Column B contains the full list of suggestions submitted by the group for discussion.*

*SAY: We have a few (OR more than a few) suggestions here. Let's tackle these suggestions in two stages. First, we'll try to group the suggestions together.  Try to think of a group or category that each of these practices fits into so that we can keep similar practices close together.  If you think multiple suggestions are describing the same practice, or if you do not understand what practice is being suggested, please say so.*

***DO:***
- Type a group name for each practice after the experts have come close to a consensus. This group name should go in the same cell, formatted as 'group name: practice.'
- If everyone agrees that a practice has been suggested multiple times, you may delete all but one of them (by deleting the contents of the cells in column A.)
- Type any comments that you would like to save but not share with the experts in column B.
- Revise the wording of certain practices if experts agree that it is ambiguous.

***DO:*** Once all the suggestions have been addressed, click 'Round Two: Update Suggestion List.' This will remove empty rows (from deleted practices) and re-alphabetize the list to move groups together.  The suggestion list will also be updated for the experts.

*SAY: Notice that the list of practices in your column B has updated.  Now that the practices are categorized, we should evaluate the wording of each of the practices.  The goal here is to make the description of each practice objective and unambiguous.*

***DO:*** After you have changed the wording of a practice, you may click 'Round Two: Update Suggestion List' to reorganize your list and also update the expert's copy of the list.

<u>Round Three</u>

*SAY: Please switch to the Round Three worksheet. Now, we would like you to evaluate a list of proposed policies.  The criteria are: "H" have tried it, "I" for important, "N" for not important, "F" for feasible.  For each parameter which applies, enter an 'X' in that cell.*

<u>Round Three Expert Input</u>

***DO:*** Make sure you are starting to see responses appear in your *Round Three* tab. Once all participants are finished entering their judgements, it is time to discuss the (anonymized) results with them.

*SAY*: *Now we are going to view the tallied results.*

***DO:*** Find and click 'Round Three: Tally Results' in the menu bar, as shown below. The tallied judgements will be added to the rightmost columns of your Round One sheet.

<u>Round Three Discussion</u>

*SAY*: *You should now all be able to see the tallied results in your Round Three sheet.*

*SAY*: *Discussion similar to Round One*

<u>(Optional) Round Four</u>

*SAY*: *We have the option of evaluating the list of suggestions from Round Two. Are you willing and available for this?*

***DO:*** If there is time and the experts are willing, you may initiate the third round by finding and selecting 'Round Four: Create Worksheets.' This creates a new tab for the mastersheet and all the expert sheets. Go to the Round Four worksheet tab that has been created.

*SAY*: *You should now select the Round Three worksheet tab. We'll follow the same protocol as Round One. Go through each practice and select which parameters apply.*

***DO:*** Follow the instructions for Round One (immediately after the Round One Begins heading.)

<u>Final Comments</u>

*SAY*: *Thank you again for your input. We expect to use your input, together with the important new suggestions that you have generated, in our further discussions with other security practitioners. We will be sure to send you a copy of our project report. In fact, we may even ask you to comment on it before it is finalized. Please do follow up via email or phone with any feedback or additional ideas you may have.*